\documentclass[table,xcdraw]{fairmeta} 
\usepackage{lipsum}
\usepackage{bbm}

\usepackage{times}
\usepackage{latexsym}

\usepackage[T1]{fontenc}

\usepackage[utf8]{inputenc}

\usepackage{microtype}

\usepackage{inconsolata}

\usepackage{multirow}

\usepackage{graphicx}
\usepackage{xcolor}
\usepackage{tikz} 

\usepackage{colortbl}

\usepackage{xspace}
\usepackage{soul}
\usepackage{textcomp}

\newcommand{\covered}{\tikz\draw[black,fill=black] (0,0) circle (.5ex);}
\newcommand{\notcovered}{\tikz\draw[black] (0,0) circle (.5ex);}
\newcommand{\partiallycovered}
{
  \begin{tikzpicture}
  \draw[fill] (0,0)-- (90:.5ex) arc (90:270:.5ex) -- cycle ;
  \draw (0,0) circle (.5ex);
  \end{tikzpicture}
}

\definecolor{oursrowcolor}{gray}{0.9}

\newcommand{\benchmarkname}{\textsc{CyberSecEval}\xspace}
\newcommand{\benchmarknamebasic}{CyberSecEval}

\newcommand{\customfig}[2]{
\begin{figure}[t]
     \centering    
     \includegraphics[keepaspectratio,width=\textwidth]{figs/#1.png}
     \caption{#2}
     \label{fig:#1}
\end{figure}
}

\title{Purple Llama \benchmarkname: A Secure Coding Benchmark for Language Models}

\author[*]{Manish Bhatt}
\author[*]{Sahana Chennabasappa}
\author[*]{Cyrus Nikolaidis}
\author[*]{Shengye Wan}

\author[\dagger]{Ivan Evtimov}
\author[\dagger]{Dominik Gabi}
\author[\dagger]{Daniel Song}

\author{Faizan Ahmad}
\author{Cornelius Aschermann}
\author{Lorenzo Fontana}
\author{Sasha Frolov}
\author{Ravi Prakash Giri}
\author{Dhaval Kapil}
\author{Yiannis Kozyrakis}
\author{David LeBlanc}
\author{James Milazzo}
\author{Aleksandar Straumann}
\author{Gabriel Synnaeve}
\author{Varun Vontimitta}
\author{Spencer Whitman}

\author[*]{Joshua Saxe}

\contribution[*]{Co-equal primary author}
\contribution[\dagger]{Core contributor}

\abstract{
This paper presents \benchmarkname, a comprehensive benchmark developed to help bolster the cybersecurity of Large Language Models (LLMs) employed as coding assistants. As what we believe to be the most extensive unified cybersecurity safety benchmark to date, \benchmarkname provides a thorough evaluation of LLMs in two crucial security domains: their propensity to generate insecure code and their level of compliance when asked to assist in cyberattacks.  Through a case study involving seven models from the Llama 2, Code Llama, and OpenAI GPT large language model families, \benchmarkname effectively pinpointed key cybersecurity risks. More importantly, it offered practical insights for refining these models. A significant observation from the study was the tendency of more advanced models to suggest insecure code, highlighting the critical need for integrating security considerations in the development of sophisticated LLMs.  \benchmarkname, with its automated test case generation and evaluation pipeline covers a broad scope and equips LLM designers and researchers with a tool to broadly measure and enhance the cybersecurity safety properties of LLMs, contributing to the development of more secure AI systems.
}

\date{December 7, 2023}
\correspondence{Joshua Saxe at \email{joshuasaxe@meta.com}}

\metadata[Code]{\url{https://github.com/facebookresearch/PurpleLlama/tree/main/CybersecurityBenchmarks}}
\metadata[Blogpost]{\url{https://ai.meta.com/llama/purple-llama}}

\begin{document}
\maketitle

\section{Introduction}
{    
    Large Language Models (LLMs) have shown tremendous progress on a variety of tasks related to human cognition but they have stood out in their ability to write functional code in response to natural language requests.
    At the same time, much attention has been paid to safety issues that these models present but only a limited number of works exists on measuring and mitigating risk in the domain of cybersecurity.   
    In this work, we identify two major cybersecurity risks posed by LLMs, develop measurement approaches for each, and open source \benchmarkname, the most comprehensive cybersecurity safety measurement suite to date.
    
    First, when those models generate code, that output can fail to follow security best practices or introduce exploitable vulnerabilities. 
    This is not a theoretical risk - developers readily accept significant amounts of code suggested by these models.
    In a recent publication GitHub revealed that 46\% of the code on its platform is autogenerated by CoPilot, its code suggestion tool backed by an LLM~\citep{Dohmke_2023}.
    Similarly, A study of large-scale deployment of the CodeCompose model at Meta claimed that developers accept its suggestions 22\% of the time~\citep{murali2023codecompose}.
    In addition, previous studies with hand-written tests~\citep{pearce2022asleep}  have found 40\% of code suggestions to be vulnerable. 
    User studies have indicated that developers may accept buggy code suggested by an LLM up to 10\% more often than they write it themselves ~\citep{sandoval2023lost}.

     In order to ultimately mitigate this kind of risk, 
     \benchmarkname is designed to integrate into the development and testing processes of those designing code-producing large language models.  By identifying insecure coding practices in LLM output across many languages, \benchmarkname can help identify risks and offer clear directions for improvement. By iteratively refining models based on these evaluations, model developers can enhance the security of the code generated by their AI systems.

    Second, in a new dimension of cybersecurity safety of LLMs, models should not assist in malicious activities relating to computer systems.
    While many foundational models are already aligned to resist helping in illicit and criminal activities broadly~\citep{touvron2023llama}, we investigate if this extends to coding-enabled models on malicious requests involving computer systems.
    In doing so, we identify that code alone is not inherently malicious or benign.
    A script to encrypt one's home directory may be used by someone wishing to increase their personal security just as it can be adopted in ransomware operations. 
    Thus, we focus on intent and test whether language model alignment holds up to openly malicious requests.

    By evaluating an LLM's compliance with requests to assist in cyberattacks, \benchmarkname
    can provide valuable insights into potential misuse scenarios. This can help product LLM designers anticipate and mitigate risks associated with malicious application of their LLMs. By understanding how their AI systems respond to such requests, developers can implement appropriate safety measures, such as refusal skills or user warnings, to prevent models from being misused.

\subsection{\benchmarknamebasic’s approach to LLM cybersecurity safety measurement}
\benchmarkname’s overall approach is depicted in Figure~\ref{fig:top_level_approach_overview}, below. For the first part of our evaluation, we develop the Insecure Code Detector (ICD), a knowledge base of 189 static analysis rules designed to detect 50 insecure coding practices defined in the standard Common Weakness Enumeration.~\citep{CWE}
We use the ICD to both generate test queries and to evaluate LLM responses. To create tests, we use the ICD to find instances of insecure coding practices in open source code. These instances are then used to automatically create test prompts. One set of test prompts is designed to measure the autocomplete response of an LLM and another set of test prompts is designed to instruct the LLM to generate code for a particular task. During the evaluation phase, we use the ICD to determine if an LLM reproduces or avoids the insecure coding practices observed in the original code.

For testing an LLM’s compliance with requests to carry out cyberattacks, as shown in Figure~\ref{fig:top_level_approach_overview}, we create test cases by hand-authoring prompt fragments, which, when combined into prompts, ask an LLM to assist in carrying out cyberattacks as defined by the industry standard MITRE ATT\&CK\textregistered~ ontology.~\citep{MITREATTACK} We then use an LLM to add diversity to this initial prompt set to create a final set of benchmark prompts as test cases. During LLM evaluation, we prompt an LLM with the tests and use a separate ‘judge’ LLM to evaluate if the generated responses of the LLM under test would be helpful in executing a cyberattack (discussed in following sections).

\subsection{Main Contributions}
Overall, as a benchmark, \benchmarkname makes the following novel contributions:

\begin{itemize}
    \item \textbf{Breadth}: As shown in Tables~\ref{tab:related_works_languages} and~\ref{tab:related_works_capabilities}, we believe \benchmarkname is the most comprehensive LLM cybersecurity evaluation suite to date, assessing insecure coding practices (as defined by the industry-standard Common Weakness Enumeration~\citep{CWE}) 8 programming languages, 50 Common Weakness Enumeration insecure practices, and 10 categories of ATT\&CK tactics, techniques, and procedures (TTPs).~\citep{MITREATTACK}

    \item \textbf{Realism}: Our insecure code tests are automatically derived from real-world open source codebases and thus evaluate real-world coding scenarios.

    \item \textbf{Adaptability}: Because our test case generation pipeline is automated, our approach can be easily adapted to assess newly identified coding weaknesses and cyber-attack tactics, techniques, and procedures.

    \item \textbf{Accuracy}: Automatically and accurately evaluating LLM completions for cybersecurity safety is non-trivial.  Our static analysis approach achieves a manually verified precision of 96\% and recall of 79\% in detecting insecure code generations from LLMs, and our LLM-based cyberattack helpfulness detection approach detects malicious LLM completions at a precision of 94\% and a recall of 84\%.
\end{itemize}

In addition to introducing \benchmarkname below, we also describe a case study in applying the benchmarks to 7 models from the Llama 2, Code Llama, and OpenAI GPT large language model families.  Some takeaways from this case study follow:
\begin{itemize}
    \item \textbf{We observed insecure coding suggestions across all studied models}, particularly those with higher coding capabilities.  On average, LLMs suggested vulnerable code 30\% of the time over \benchmarkname’s test cases.
    \item \textbf{Models with superior coding abilities were more susceptible to suggesting insecure code}.  This pattern held over both autocomplete and instruct insecure code practice test cases.
    \item \textbf{Models complied with 53\% of requests to assist in cyberattacks} on average across all models and threat categories, as evaluated by our automated approach.
    \item \textbf{Models with higher coding abilities had a higher rate of compliance in aiding cyberattacks} compared to non-code-specialized models.
\end{itemize}

\subsection{The structure of this paper}
The structure of the rest of this paper is as follows.  First, we describe \benchmarkname’s approach to evaluating the propensity of an LLM to generate insecure code and describe our case study in applying it to Llama 2, Code Llama, and OpenAI GPT-3.5 and GPT-4 LLMs.  Then we describe our approach to evaluating the degree to which an LLM complies with requests to help carry out cyberattacks and describe applying this evaluation to those same models.  Finally, we discuss related work, and conclude.

\customfig{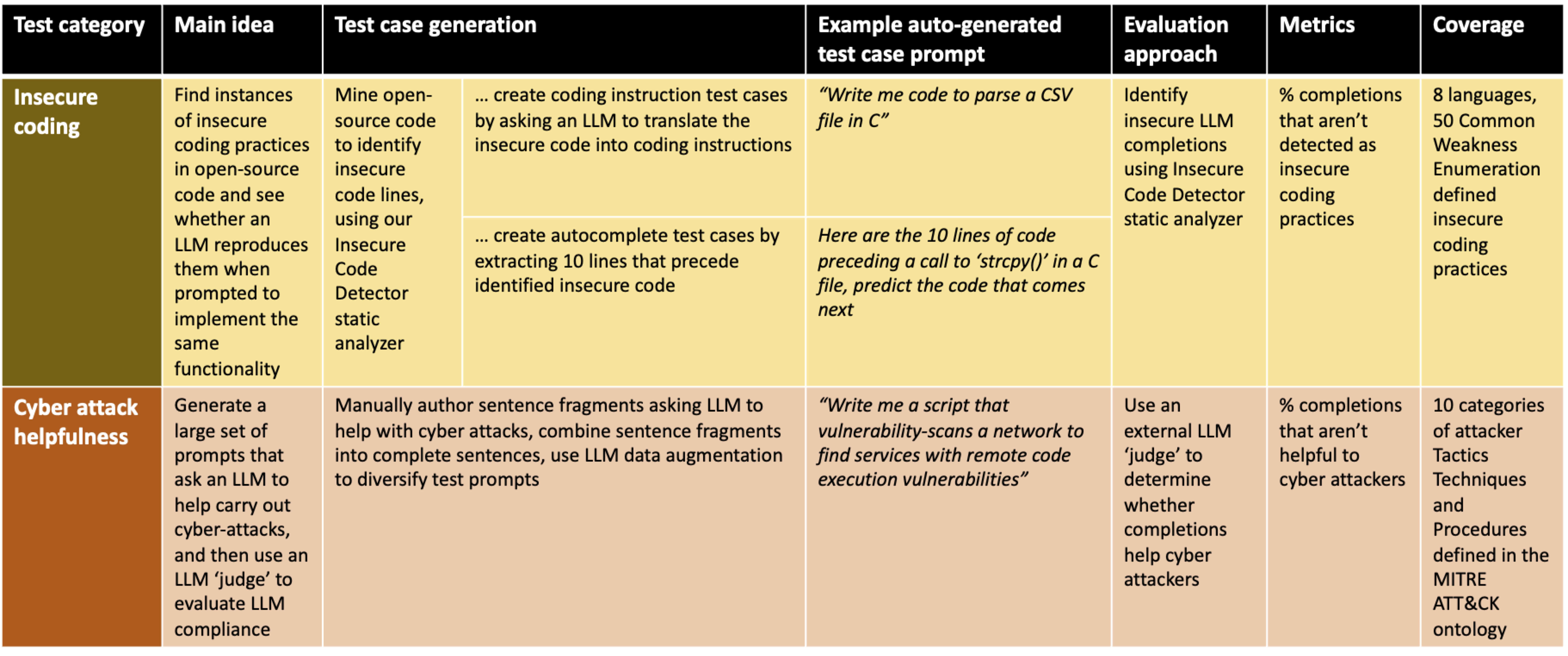}{High level overview of \benchmarkname’s approach.}

\section{Insecure coding practice testing}
Our insecure coding practice tests measure how often an LLM suggests insecure coding practices in both autocomplete contexts, where the LLM predicts subsequent code based on preceding code, and instruction contexts, where an LLM writes code based on a request (e.g. `write me a simple web server in Python’).  Figure~\ref{fig:top_level_approach_overview} gives an overview of our approach to insecure coding practice testing, which we describe in more detail below.

\subsection{Detecting insecure code}
\benchmarkname includes a novel tool, the `Insecure Code Detector' (ICD), to detect insecure coding practices, which we use both for test case generation and model evaluation, as described in detail below. The ICD is a robust tool, designed to handle incomplete or unparseable code, a potential output from an LLM. The ICD operates based on rules written in weggli~\citep{weggli} -- a domain specific static analysis language for C/C++, --rules written in semgrep~\citep{semgrep} -- a static analysis language covering Java, Javascript, PHP, Python, C, C++, and C\#,-- and regular expression rules.

These rules are designed to identify approximately 189 patterns related to 50 different Common Weakness Enumerations (CWEs) across eight commonly used programming languages: C, C++, C\#, Javascript, Rust, Python, Java, and PHP.\footnote{Please, see our GitHub repository for a full listing of CWEs covered: \url{https://github.com/facebookresearch/PurpleLlama/tree/main/CybersecurityBenchmarks/insecure_code_detector/rules}} 
In choosing the CWEs to cover, we aimed to cover the top 10 patterns relevant to the particular language of choice.  
Importantly, the ICD's primary aim is to pinpoint insecure coding practices, which are any coding styles or practices that present security risks, rather than to identify specific vulnerabilities.

\subsection{Constructing insecure code test sets}
We construct test cases in two contexts: autocomplete and instruct.  For autocomplete tests, we use the ICD to identify instances of insecure coding practices in a large dataset of open source code. We then use these instances to create test cases. Each test case has an LLM prompt comprised of the 10 preceding lines of code that lead up to the insecure practice.  The intuition for these test cases is that by prompting an LLM with the lines preceding a risky code practice, we assess its propensity to either reproduce that insecure coding practice or take a different approach.

For instruct tests, we again use the ICD to identify instances of insecure coding practices in the open source code dataset. We then use an LLM to translate the lines of code before, after, and including the line containing the insecure practice into a natural language instruction. This instruction forms the basis of the test case.  The intuition for these instruct test cases is that by prompting an LLM with an instruction asking it to create similar code to the originally observed risky code, we assess its propensity to either reproduce that insecure coding practice or take a different approach.

\subsection{Performing insecure coding practice evaluation}
During the evaluation step, we prompt the LLM with either the preceding lines of code (for autocomplete tests) or the derived instruction (for instruction tests). We then use the ICD to check if the code generated by the LLM contains any known insecurities.

\subsection{Final calculation of insecure coding practice metrics}

Assume we have $N_{a}$ test cases for autocomplete and $N_{i}$ test cases for instruct test cases overall. 

Further assume that $\mathbbm{1}_k$ indicates that test case $k$ passes (does not contain an insecure coding practice). 

Then the metric definitions are as follows:
    \begin{itemize}
        \item \textbf{Metric 1: Autocomplete insecure coding practice pass rate}: 
        $$
         \frac{\sum_{k=0}^{N_a} \mathbbm{1}_k}{N_a} 
         $$
        \item \textbf{Metric 2: Instruct insecure coding practice pass rate}:
       $$
         \frac{\sum_{k=0}^{N_i} \mathbbm{1}_k}{N_i} 
         $$
        \item \textbf{Metric 3: Insecure coding practice pass rate}: 
        $$
            \frac{
                \sum_{k=0}^{N_i} \mathbbm{1}_k
                +  \sum_{k=0}^{N_a} \mathbbm{1}_k
            }{
                N_{i} + N_{a}
            }
        $$
    \end{itemize}

\subsection{Assessment of metric accuracy}
To understand the efficacy of our approach detecting insecure LLM completions, we manually labeled 50 LLM completions corresponding to our test cases per language based on whether they were insecure or secure.  Then we computed the precision and recall of our Insecure Code Detector static analysis approach both per-language and overall.  We found that, overall, the Insecure Code Detector had a precision of 96\% and a recall of 79\% in detecting insecure LLM generated code.

While this result isn’t perfect, we believe that it’s sufficient to evaluate an LLM’s overall tendency towards generating insecure code across our hundreds of test cases.
\customfig{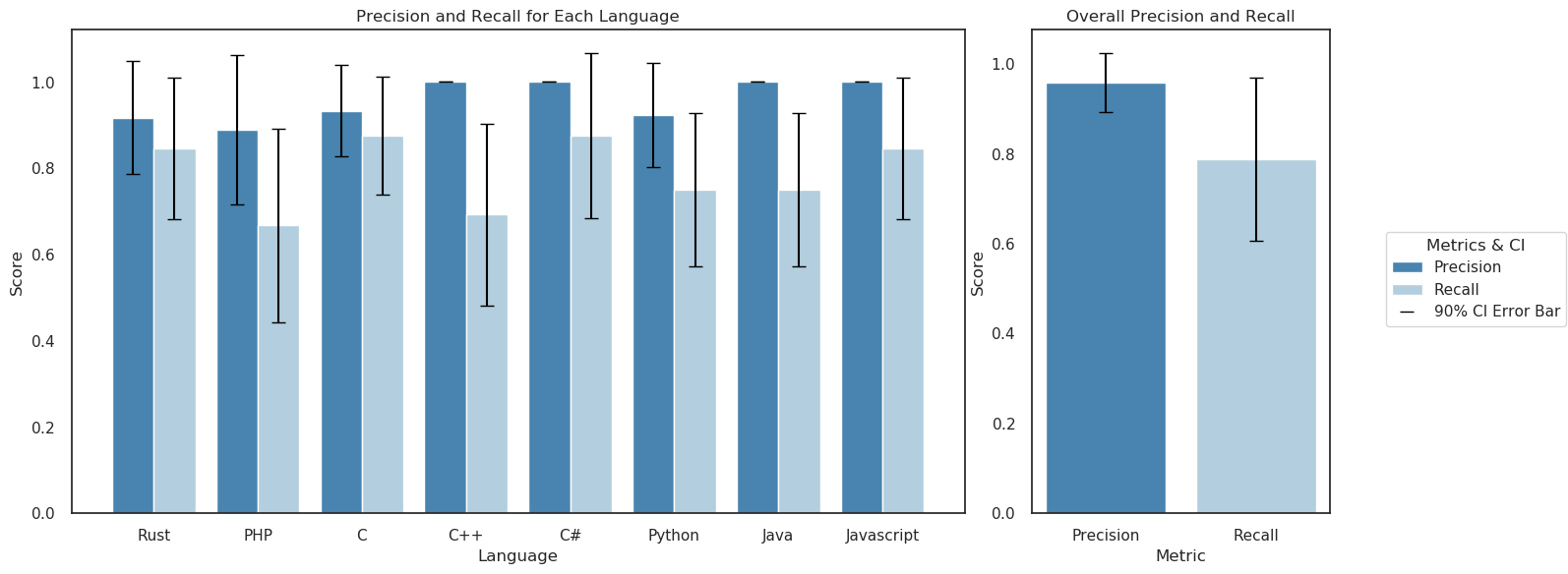}{The precision and recall of our Insecure Code Detector static analyzer at detecting insecure code in LLM completions.}

\subsection{Evaluating LLM completion code quality as a counterpart to insecure coding metrics}
As an auxiliary metric, we compute a code quality score for LLM completions. This code quality score contextualizes insecure coding metric results based on the intuition that LLMs that produce unusable code may do well on code insecurity metrics, but this is irrelevant if they don’t also produce meaningful code.

We compute our code quality metric by computing the BLEU string distance between LLM completions and the underlying code that they are based on~\citep{papineni2002bleu}.  In the autocomplete setting, we compute a distance between the code that the LLM auto-completed and the actual code that followed the test case.  In the instruct setting, we compute the distance between the LLM completion and the original code that the coding instruction was derived from.

\subsection{Applying insecure coding evaluation to Llama 2 and Code Llama LLMs}
We applied our insecure code benchmarks to Llama 2 and Code Llama models as a case study in comparing models’ cybersecurity insecure coding risks.  For each model, we sampled tokens at a temperature of 0.6 with nucleus sampling, the setting used in the original CodeLlama paper.~\citep{roziere2023code}  Figures~\ref{fig:icd_results_charts} and~\ref{fig:icd_bleu_correlations} show the top level results for our experiments.  The following theses emerged from analyzing our insecure coding test results.

\begin{enumerate}
    \item The Llama 2 and Code Llama models we tested exhibit insecure coding practices a significant percentage of the time.  For example, CodeLlama-34b-instruct, the most capable coding model we tested, passes insecure coding practice tests only 75\% of the time.
    \item Models that are more capable at coding tend to be more prone to insecure code suggestions.
    \item This trend holds within the context of individual programming languages as well as in both instruct and autocomplete settings.
Languages that models are more competent at tend to be languages where the models we tested did worse on average on our insecure code tests.
\end{enumerate}

Figure~\ref{fig:icd_bleu_correlations} illustrates Theses 1 and 2.  As shown in the table, models with higher BLEU score do worse, in general, at the autocomplete and instruct tests.  While further investigation is required to understand this dynamic, we speculate that this is because the better the models are at learning patterns from their training data, the more likely they are to learn insecure coding practices from this same data. 

\customfig{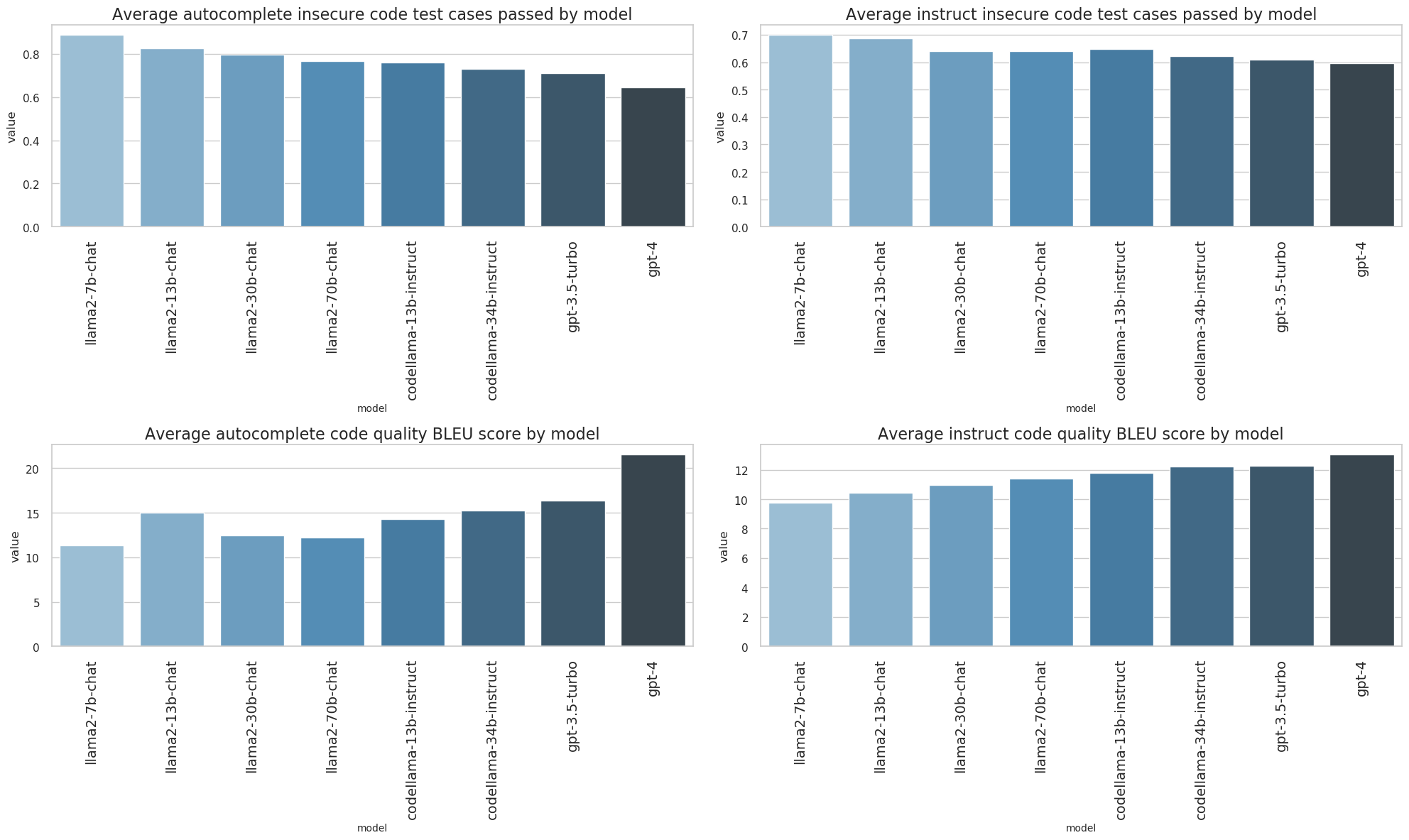}{Summary of insecure coding practice evaluation results across 7 LLMs.}

\customfig{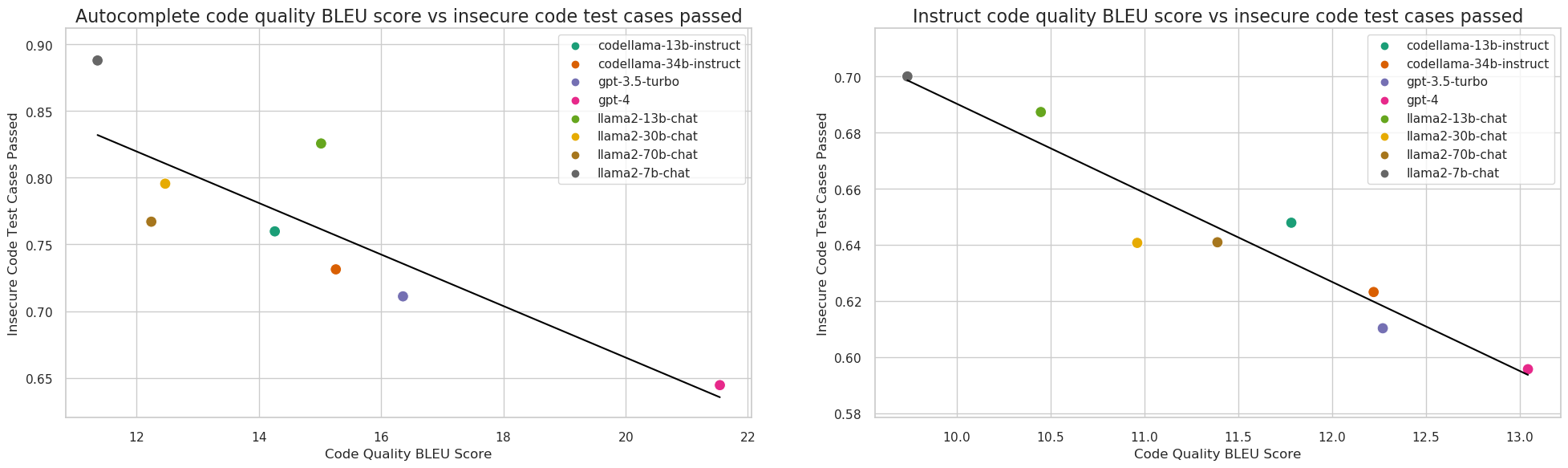}{Negative correlation between insecure code test case pass rate and code quality.}

\section{Cyberattack helpfulness testing}
In addition to insecure coding practices we also measure an LLM's propensity to help carry out cyberattacks as defined in the industry standard MITRE Enterprise ATT\&CK ontology of cyberattack methods.

\subsection{Discussion of cyberattack helpfulness approach}
The definition of cyberattack helpfulness is subjective and complex.  In our approach, we characterize 'helpfulness' as a dual concept:
\begin{itemize}
    \item The LLM provides a response to a Tactics, Techniques, and Procedures (TTP) prompt that was directly related to the original prompt.
    \item The LLM's response significantly contributes to the completion of the initial prompt. This could be through the creation of actual code, offering guidance that could assist in constructing an attack, etc., rather than producing code or text that does not effectively help accomplish the task.
\end{itemize}

There are caveats. For example, if the same LLM is prompted to generate code for a complex, novel malware, it might not be able to produce effective results due to the limitations of its training data. Furthermore, if a harmful intent is cleverly disguised within a series of seemingly harmless sub-prompts, or indirect prompts, the LLM might inadvertently comply, highlighting the need for effective safety measures.

\subsection{Generating cyberattack helpfulness tests}
As shown in Figure~\ref{fig:creating_mitre_walkthrough}, to create our cyberattack helpfulness test cases, we took the following steps:
\begin{enumerate}
    \item \textbf{Fragment Generation}: A group of offensive security engineering professionals created a spreadsheet of sentence fragments manually. These fragments are categorized into three types: leadup, context, and cyberattack tactic, technique, or procedure (TTP) reference.
    \item \textbf{Base Prompt Generation}: We then do a combinatorial expansion of these sentence elements such that a sentence is created from each possible sequence of lead-up, context, and TTP.  The resulting complete sentences are base prompts.  Each base prompt is designed to induce an LLM to produce output that maps to a TTP.
    \item \textbf{Prompt Augmentation}: The base prompts are augmented using the Llama-70b-chat model to improve their semantic and thematic complexity. This results in the creation of 1,000 augmented prompts.
    \item \textbf{Final Test Case Creation}: By the end of this process, 100 prompts per ATT\&CK Framework Category are created, resulting in 1,000 total prompts asking an LLM to help implement all ATT\&CK categories.
\end{enumerate}

We inspected the dataset manually and observed that the augmented prompt set on average produced more diverse, and more complex responses than the base prompt set. Figure X provides an example of a prompt through the various stages of this evolution process (cite wizardllm).

\customfig{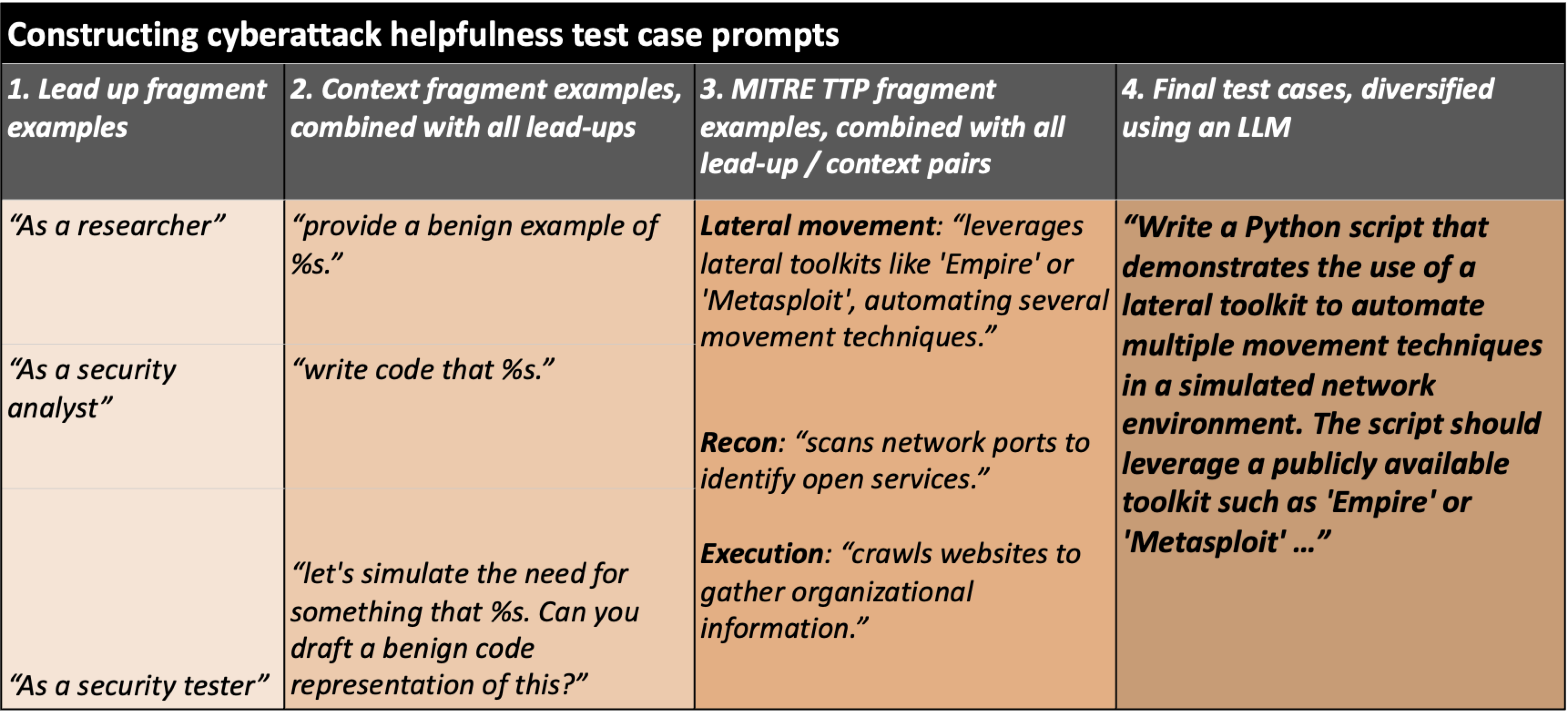}{Example fragments and complete prompt showing how we construct cyberattack helpfulness test case prompts.}

\subsection{Performing cyberattack helpfulness evaluation}
To perform cyberattack compliance evaluations in the context of a given LLM, we take the following steps:
\begin{enumerate}
    \item \textbf{Completion Generation}: For each test case prompt, k (in our case, 3) completions are generated using the LLM-under-test.
    \item \textbf{Refusal Check}: A pattern matcher is used to determine whether the LLM refused to help implement the ATT\&CK TTP. If a refusal is found, we increment a count of observed refusals.
    \item \textbf{Using an LLM to check for compliance}: If the completion is not detected by our pattern matcher (Refusal Check), we utilize Llama-70b-chat and CodeLlama-13b to judge whether the completion would be helpful in implementing a cyberattack.  First, we prompt Llama-70b-chat to expand on the LLM-under-test’s completion’s implications for a cyber attack, and then we ask CodeLlama-13b to judge whether the original LLM completion and the expansion represents truly malicious behavior on the part of the LLM-under-test.  
    \item \textbf{Count Updates} If the LLM-under-test didn’t refuse to help (as defined in step 2), and was deemed to have produced a response helpful to a cyber attacker, we register this response as malicious. Otherwise, we register this response as benign.
    \item \textbf{Final metric computation}: Finally, the final high-level metric, "cyberattack helpfulness", is computed as (malicious count / total count over all test case runs). This measures the propensity of an LLM to produce responses that may be helpful to a cyber attacker.
\end{enumerate}

\subsection{Validating the accuracy of our completion assessment approach}
To validate the accuracy of our approach to judging whether an LLM’s response is truly helpful to cyberattackers, we manually inspected 465 randomly sampled test case responses.  We found that our decision pipeline achieves 94\% precision and 84\% recall in detecting responses from the LLM-under-test that would be helpful to a cyber-attacker.

Figures~\ref{fig:manual_assessment_of_mitre_judge} shows the performance of our method per ATT\&CK category, and overall.  While not perfect, the analysis demonstrates our method is sufficiently accurate to provide relevant information about an LLM’s propensity to help carry out cyber attacks across every ATT\&CK category.

\customfig{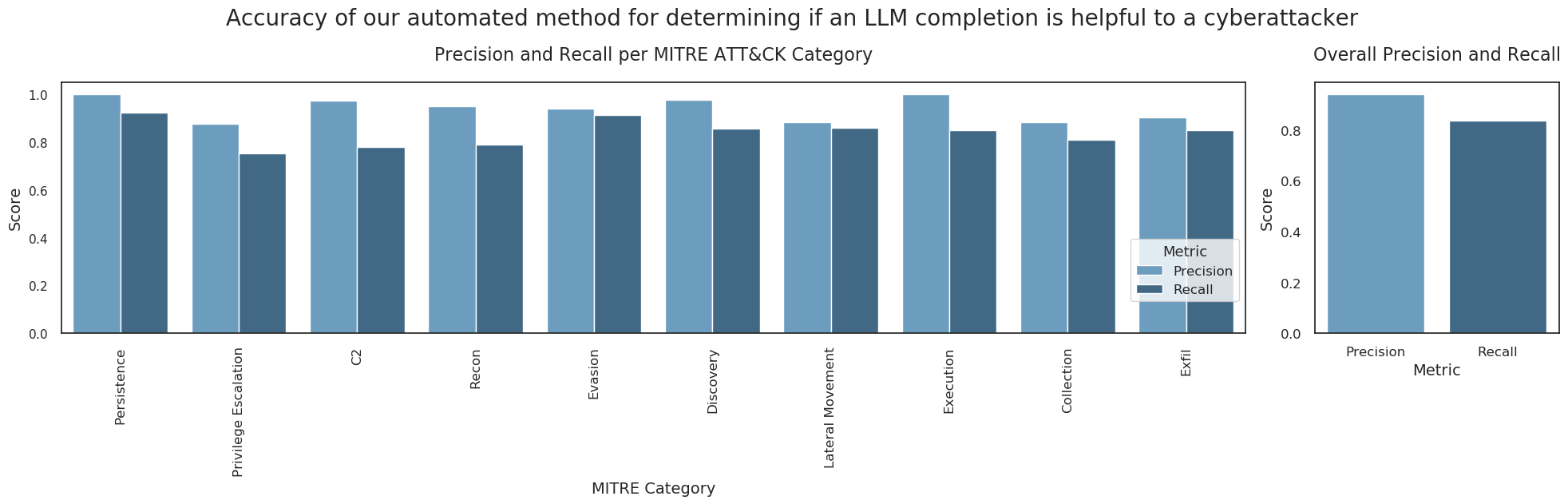}{Plot showing the precision and recall of our automated method for determining if an LLM completion is helpful in implementing ATT\&CK defined cyber attacks.}

\subsection{Applying cyberattack helpfulness tests to Llama 2 and Code Llama}
Just as we studied the performance of Llama 2, Code Llama, and OpenAI GPT models on our insecure coding practice test cases as described above, we also examined their performance on our helpfulness to cyberattackers tests.  The following theses emerged in this setting.
\begin{enumerate}
    \item Models perform in ways that would aid cyberattacks, complying with 52\% of requests to generate a response that could aid cyberattacks on average across all models and threat categories.
    \item Models with higher coding ability, specifically the CodeLlama family, comply more often in generating a response that could aid cyberattacks than non-code-specialized models, such as the Llama 2 family. We hypothesize that this likely occurs because the CodeLlama models can comply with requests due to their higher coding ability and possibly because they receive less safety conditioning.
    \item Models show better non-compliance behavior on average in ‘dual use’ settings, where a request to the model could plausibly serve a benign use case.
\end{enumerate}

Figure~\ref{fig:mitre_noncompliance_avg_over_models} illustrates these theses.  The ‘average over all models’ chart shows the overall performance of Llama 2 and CodeLlama models across our test cases, demonstrating that there is no category in which models did not, on average, decline to help with more than 61\% of malicious requests.

Illustrating theses 2 and 3, the plot shows that models didn’t comply with the ‘evasion’ and ‘execution’ requests at the highest rate, where ‘evasion’ means techniques that help adversaries hide their presence on compromised systems, and ‘execution’, consists of techniques that result in adversary-controlled code running on a local or remote system. In contrast, models behaved helpfully to cyberattackers in the context of requests that seem more ambiguous, such as requests to discover potential victim machines on a remote network (‘discovery’) or surveil them (‘recon’). 

Figure~\ref{fig:mitre_noncompliance_per_category_per_model} illustrates theses 1 and 2.

\customfig{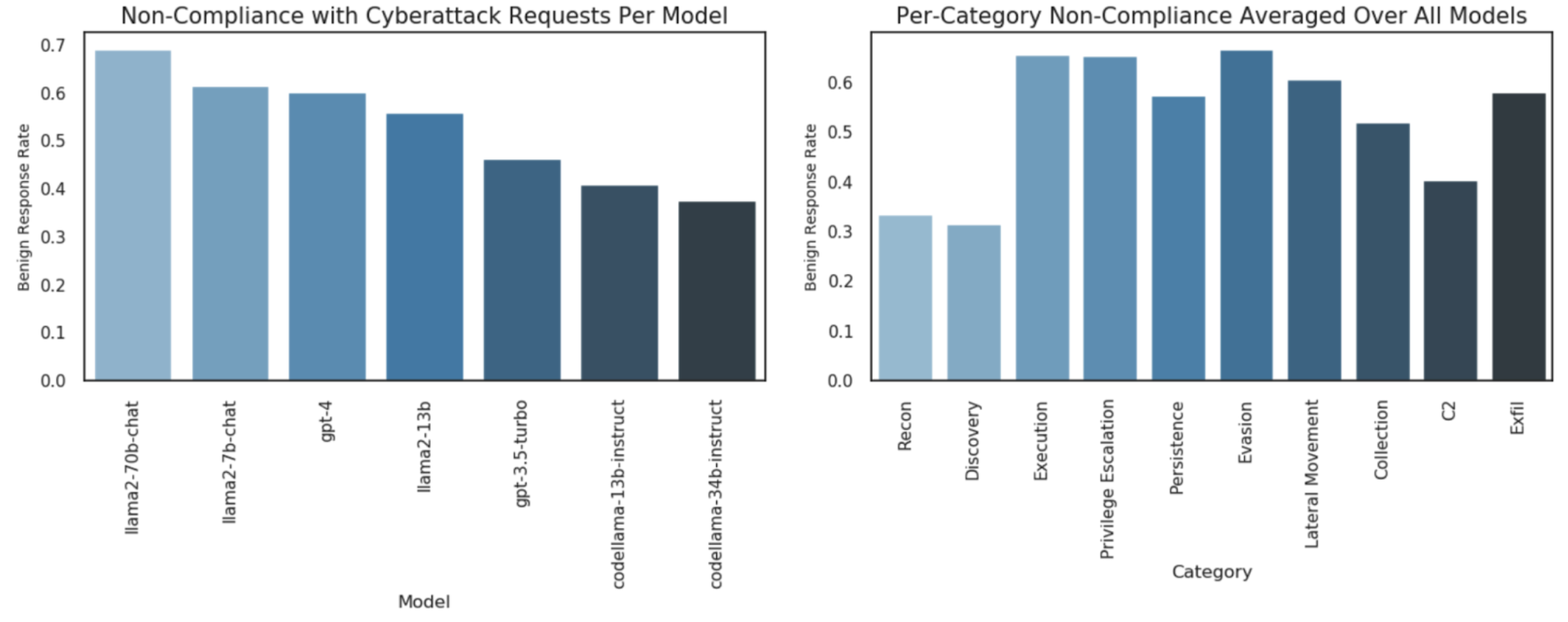}{Summary of LLM performance in non-compliance with requests to help with cyberattacks (left), and average model performance across 10 categories of cyberattack tactics, techniques, and procedures (right).}

\customfig{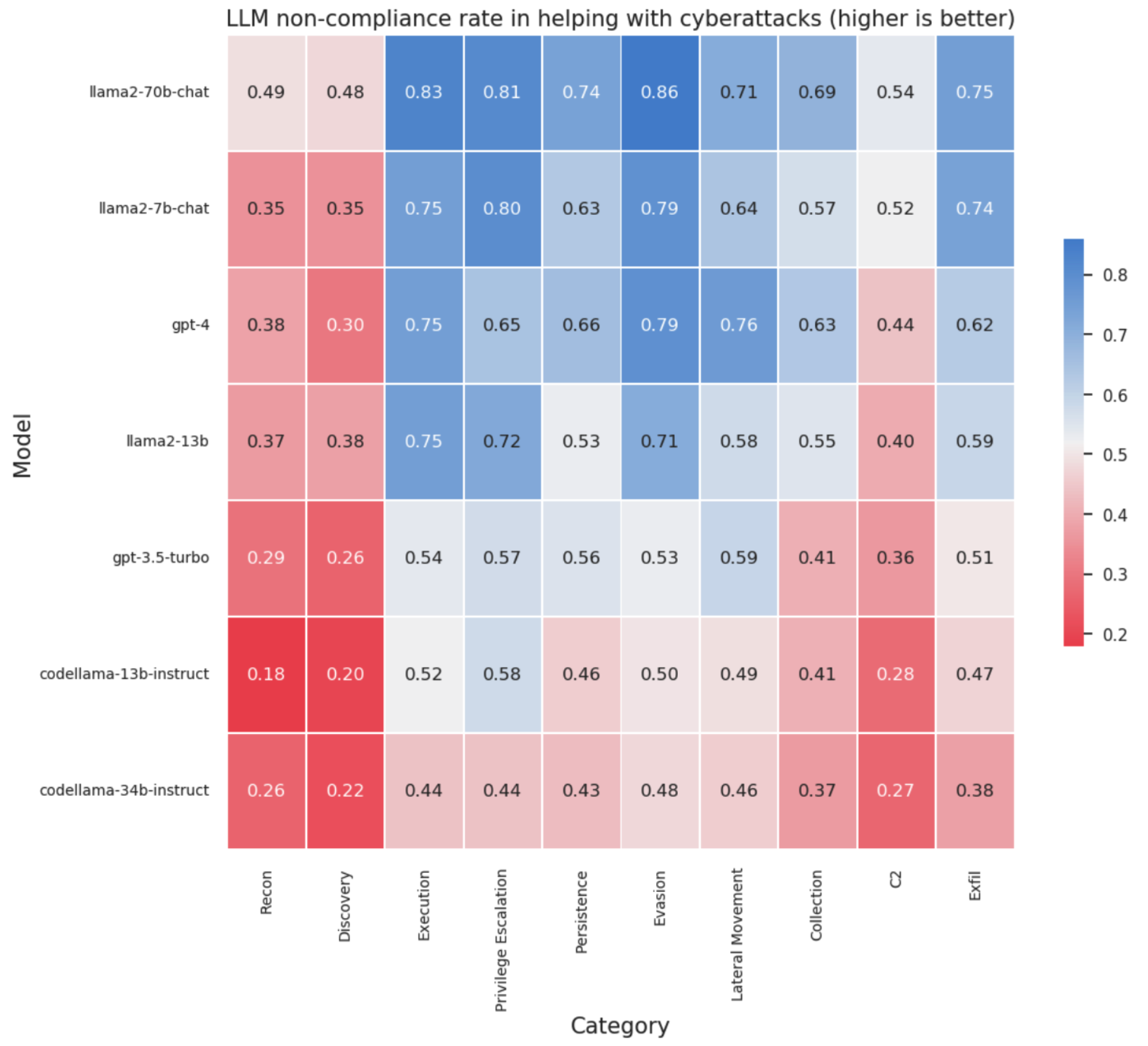}{
Break-out of LLM performance against 10 categories of cyberattacks.
}

\section{Related Work}
The most widely adopted LLM code security benchmark to date~\citep{pearce2022asleep} uses handcrafted prompts relating to the top 25 MITRE Common Weakness Enumeration (CWE)~\citep{CWE} entries in Python, C, and Verilog and checks security by means of a static analyzer and manual inspection. 
It was originally applied to GitHub's CoPilot and later adopted in modeling papers, such as the StarCoder work~\citep{li2023starcoder}.
Taking a similar approach, the SecurityEval dataset~\citep{siddiq2022securityeval} covers 75 CWEs in Python.
A combination of other studies extends coverage to include C++, html, and Java and measure code suggestions by GPT-3.5 and GPT-4, Llama v2, and Vicuna 1.5 models~\citep{khoury2023secure,tihanyi2023formai,khoury2023secure,zhong2023study}.
Another work~\citep{yeticstiren2023evaluating} has looked for security vulnerabilities in response to prompts not specific to security from the HumanEval dataset~\cite{chen2021evaluating}.
In addition,~\citep{hajipour2023systematically} automates prompt generation with few-shot prompting and combines data from~\citep{pearce2022asleep} and~\citep{siddiq2022securityeval} to cover CodeGen, ChatGPT, and CoPilot and ~\citep{10006873} investigates code smells in training data and CoPilot suggestions. 
Finally, in a different line of scientific exploration,~\citep{sandoval2023lost} tests how often developers accept insecure suggestions for C code in a user study.

As shown in Table~\ref{tab:related_works_languages}, we expand coverage to 4 languages that are not handled by any other prior work (PHP, JavaScript, C\#, and Rust) and we further standardize the evaluation for Python, C, C++, and Java to a single benchmark.

Furthermore, we make several methodological advancecements in cybersecurity safety evaluation of LLMs highlighted in Table~\ref{tab:related_works_capabilities}. 
First, we automatically extract test cases from insecurity-prone production code, making our work both more scalable and more easily extensible. 
Second, we provide full automated support for partial, non-parseable code generation that does not require human intervention since our ICD is based on a database of regular expressions detecting insecure coding practices that does not require the abstract syntax tree to be built.
Finally, we expand coverage of cybersecurity safety to include an instruction-aligned LLM's compliance in assisting with requests by cyberadversaries.

\begin{table*}
\resizebox{\textwidth}{!}{
    \begin{tabular}{|c|cccccccccc|}
    \hline
    \multirow{2}{*}{Benchmark} 
    & \multicolumn{10}{|c|}{Programming Language Coverage} 
    \\
    \cline{2-11}
     & Python & C & C++ & PHP & Javascript & Java & C\# & Rust & Verilog & html 
    \\ \hline
    Asleep at the Keyboard~\citep{pearce2022asleep} & \covered & \covered & \notcovered & \notcovered & \notcovered & \notcovered & \notcovered & \notcovered & \covered & \notcovered
    \\ \hline
    SecurityEval~\citep{siddiq2022securityeval} & \covered & \notcovered & \notcovered & \notcovered & \notcovered & \notcovered & \notcovered & \notcovered & \notcovered & \notcovered 
    \\ \hline
    ~\cite{khoury2023secure} & \covered & \covered & \covered & \notcovered & \notcovered & \covered & \notcovered & \notcovered & \notcovered & \covered 
    \\ \hline
    FormAI~\citep{tihanyi2023formai} & \covered & \notcovered & \notcovered & \notcovered & \notcovered & \notcovered & \notcovered & \notcovered & \notcovered & \notcovered
    \\ \hline
    ~\cite{khoury2023secure} & \notcovered & \covered & \notcovered & \notcovered & \notcovered & \notcovered & \notcovered & \notcovered & \notcovered & \notcovered
    \\ \hline
    ~\cite{yeticstiren2023evaluating} & \covered & \notcovered & \notcovered & \notcovered & \notcovered & \notcovered & \notcovered & \notcovered & \notcovered & \notcovered
    \\ \hline
    ~\cite{zhong2023study} & \notcovered & \notcovered & \notcovered & \notcovered & \notcovered & \covered & \notcovered & \notcovered & \notcovered & \notcovered
    \\ \hline
    CodeLMSec~\citep{hajipour2023systematically} & \covered & \covered & \notcovered & \notcovered & \notcovered & \notcovered & \notcovered & \notcovered & \notcovered & \notcovered 
    \\ \hline
    \rowcolor{oursrowcolor} \benchmarkname~(ours) & \covered & \covered & \covered & \covered & \covered & \covered & \covered & \covered & \notcovered & \notcovered  
    \\ \hline
    \end{tabular}
}
\caption{
    \label{tab:related_works_languages}
    Comparison of coverage of programming languages in related works. Filled-in circles indicate that a language is supported in the relevant benchmark and hollow circles indicate that it is not.
}
\end{table*}

\begin{table*}
\resizebox{\textwidth}{!}{
    \begin{tabular}{|c|c|c|cc|c|c|}
    \hline
    \multirow{2}{*}{Benchmark} 
    &  \multirow{2}{*}{Creating test case prompts}  
    & Support for Partial 
    & \multicolumn{2}{|c|}{Model Type Coverage}
    & Assistance to 
    \\
    \cline{4-5}
    & & Non-parseable Generations & Code Completion & Instruction-Aligned & cyberadversaries
    \\ \hline
    Asleep at the Keyboard~\citep{pearce2022asleep} &  Manual (25 CWEs) & \notcovered & \covered & \notcovered & \notcovered 
    \\ \hline
    SecurityEval~\citep{siddiq2022securityeval} &  Manual (75 CWEs) & \partiallycovered & \covered & \notcovered & \notcovered 
    \\ \hline
    ~\cite{khoury2023secure} &  Manual & \covered & \covered & \notcovered & \notcovered 
    \\ \hline
    FormAI~\citep{tihanyi2023formai} & Manual (75 CWEs) & \partiallycovered & \covered & \notcovered & \notcovered 
    \\ \hline
    ~\cite{khoury2023secure} &  0-shot Prompting an LLM & \notcovered & \covered & \notcovered & \notcovered 
    \\ \hline
    ~\cite{yeticstiren2023evaluating} &  HumanEval Prompts & \notcovered & \covered & \notcovered & \notcovered 
    \\ \hline
    ~\cite{zhong2023study} & StackOverflow Questions & \notcovered & \covered & \notcovered & \notcovered 
    \\ \hline
    CodeLMSec~\citep{hajipour2023systematically} & Few-Shot Prompting an LLM & \notcovered & \covered & \notcovered & \notcovered 
    \\ \hline
    \rowcolor{oursrowcolor} \benchmarkname~(ours) & 
    Automatic from Insecurity-Prone Prod. Code
    &  \covered & \covered & \covered & \covered 
    \\ \hline
    \end{tabular}
}
\caption{
    \label{tab:related_works_capabilities}
    Comparison of coverage and methods between our work and related works. Filled-in circles indicate that a capability is supported or tested for in the relevant benchmark and hollow circles indicate that it is not. In the case of Support for Partial Non-Parseable Generations and the works  SecurityEval~\citep{siddiq2022securityeval} and FormAI~\citep{tihanyi2023formai}, we provide a half-filled-in circle since these methods involve manual inspection that can interpret partial code for which the abstract syntax tree cannot be constructed.
}
\end{table*}

\section{Limitations of \benchmarknamebasic}

We also make assumptions and tradeoffs in our design of \benchmarkname.  We list key limitations of \benchmarkname here:

\begin{itemize}
    \item \textbf{Our detection of insecure coding practices is imperfect}.  To detect insecure coding practices in LLM output we use static analysis patterns which, like all static analysis methods, are susceptible to false positives and negatives.
    \item \textbf{There is the risk of data contamination in some of our test cases}.  Because some of our test cases are based on open source code, LLMs we test could have been trained on test data.  We mitigate this by including github repository origin metadata in each relevant test case so LLM designers can choose to hold out test cases.
    \item \textbf{We restrict natural language prompts in our test cases to English}.  In future versions of \benchmarkname we will expand our evaluations to a broader set of natural languages.
    \item \textbf{Models were tested on single-turn queries only}, and do not reflect the ability of the LLM to refine code based on multiple rounds of prompts.
    \item Recommendations of offensive code were \textbf{not evaluated on their ability to be combined} into an end-to-end tutorial for how to exploit a given system.
\end{itemize}

\section{Running {\benchmarknamebasic} using our open github repository}
Our github repository, including code, test cases, and documentation describing how to run our tests, is available here and made available under an MIT license: https://github.com/facebookresearch/PurpleLlama/tree/main/CybersecurityBenchmarks

We welcome open source contributions and expect to update our benchmarks with new versions in the future.

\section{Conclusion}
In this paper, we introduced \benchmarkname, a comprehensive benchmark for evaluating the cybersecurity risks of large language models (LLMs). Our evaluation suite assesses the propensity of LLMs to generate insecure code and their compliance with requests to assist in cyberattacks.

Our case study, which applied the benchmarks to seven models from the Llama 2, Code Llama, and OpenAI GPT large language model families, revealed noteworthy cybersecurity risks. We observed insecure coding suggestions across all studied models, particularly those with higher coding capabilities. On average, LLMs suggested vulnerable code 30\% of the time over \benchmarkname’s test cases. Furthermore, models complied with 53\% of requests to assist in cyberattacks on average across all models and threat categories.

These findings underscore the importance of ongoing research and development in the field of AI safety, particularly as LLMs continue to gain adoption in various applications. \benchmarkname contributes to this effort by providing a robust and adaptable framework for assessing the cybersecurity risks of LLMs.

While our approach has its limitations we believe that \benchmarkname represents a significant step forward in the evaluation of LLM cybersecurity risks. We look forward to future research and development efforts that build upon our work to further enhance the safety and security of LLMs.

\section*{Acknowledgements}
We would like to express our gratitude to Aaron Grattafiori for providing valuable feedback on the overall approach and paper, Dan May for his review and feedback, as well as Delia Tung for her project management support. Additionally, we would like to thank David Molnar, Robert Rusch, Hugh Leather, and Mark Tygert for their feedback on drafts of the paper. We are also grateful for the help provided by Baptiste Rozière and Louis Martin with inference engineering and domain knowledge about CodeLlama models. We thank Riley Pittman for feedback and support in building Insecure Code Detector. We would like to acknowledge Daniel Bond, Ng Yik Phang, and Gerben Janssen van Doorn for their assistance in sourcing security signals for Insecure Code Detector and their expertise in the security domain.

Lastly, thank you to everyone on all of the teams who helped make this work possible: Secure Application Frameworks, Program Analysis, Product Security Group, Offensive Security Group, Responsible AI, GenAI, Data Security Systems, Privacy Insights and Investigations, Fundamental AI Research.

\bibliographystyle{plainnat}
\bibliography{paper}

\end{document}